\documentstyle[pra,psfig,aps]{revtex}

\begin{document}
\draft

\title{Quantum Revivals in a Periodically Driven Gravitational Cavity}

\author{F. Saif,\, G. Alber,\, V. Savichev,\, W. P. Schleich}
\address{Abteilung f\"ur Quantenphysik, Universit\"at Ulm,
         Albert-Einstein-Allee 11, D-89069 Ulm, Germany}
\maketitle
\begin{abstract}
Quantum revivals are investigated for the dynamics of an atom
in a driven gravitational cavity. 
It is demonstrated that the external driving field
influences the revival time significantly.
Analytical expressions are presented which are based 
on second order perturbation theory and semiclassical secular 
theory. These analytical results explain the dependence of the revival
time on the characteristic parameters of the problem 
quantitatively in a simple way. They are in excellent 
agreement with numerical results.
\end{abstract}

Periodically driven quantum systems
have received considerable attention over the past
few years due to the presence of Anderson-like 
localization~\cite{kn:benv1}.
Some recent numerical work indicates the presence of 
quantum revivals~\cite{kn:hogg} in periodically driven quantum
systems~\cite{kn:haak,kn:bres} as well as in other two-degree-of-freedom
systems~\cite{kn:toms}. Though many aspects of quantum localization are well understood by now,
so far even the most elementary questions concerning 
revival effects in periodically driven quantum systems 
are not yet comprehensible. Under which circumstances do such revival
phenomena appear and how do these
revival effects depend on the frequency and on the driving amplitude of
the applied periodic force? 
The main intention of this paper is to address these questions 
by considering a physical system which is of 
particular interest in the field of 
atom optics~\cite{kn:mlyn}, namely an atom in a periodically
driven gravitational cavity~\cite{kn:wallis}.
It will be shown that quantum revivals can be 
observed in this physical system and that the dominant influence
of the periodic driving force results in a change of the revival time.
Simple analytical results are presented which explain the
quantitative dependence of the revival time on the driving frequency 
and the strength of the driving amplitude. These analytical
results which are based on semiclassical second order 
perturbation theory and on semiclassical secular perturbation theory
are in excellent agreement with numerical results.

Let us consider an atom in a periodically
driven gravitational cavity~\cite{kn:wallis}.
The atom moves under the influence of gravity 
in the positive $\tilde z$-direction 
and is reflected as it hits a mirror. The atomic mirror is assumed to be 
made up of an evanescent electromagnetic field with electric field
strength ${\it{\bf E}}(\tilde z, t) = {\bf e}_x {\it E}_0 e^{-\omega_L \tilde z/c}
e^{-i\omega_L  t} + {\rm c.c}$.
In order to study the effect of an external driving force
on this elementary quantum system the evanescent wave is assumed to
be modulated, e.g. by an acousto-optical modulator~\cite{kn:sten},
that is  ${\it E}_0={\cal E}_0\exp(\epsilon \sin\omega t)$.
Assuming that the optical driving frequency $\omega$ is well detuned
from any atomic resonance and taking into account the symmetry of the
problem in the $(x,y)$-plane the effective one-dimensional
center-of-mass motion of the atom is governed 
by the Hamiltonian~\cite{kn:saif1}
\begin{equation}
H=\frac{{p}^2}{2M} + Mg{\tilde z}
+ \frac{\hbar \Omega_{eff}}{4} e^{-2 \omega_L{\tilde z}/c+\epsilon\sin\omega t}.
\label{ham}
\end{equation}
Here ${p}$ is the atomic center-of-mass momentum, 
$M$ is the atomic mass and $g$ denotes the gravitational
acceleration. The effective Rabi frequency $\Omega_{eff}$ 
characterizes the strength of the influence of the applied electric 
field.

The potential generated by the gravitational acceleration
and by the evanescent laser field has the
approximate form of a time dependent triangular potential
well like in the Fermi accelerator~\cite{kn:fermi}.
Thus, in the subsequent, approximate
treatment we replace this potential  by the idealized, simpler form
of a triangular well with an infinitely high potential barrier at
the position $\tilde z=\frac{c\epsilon}{2\omega_L} \sin\omega t$ 
of the mirror.
In this approximation our results become independent of details
of the evanescent laser field, such as the laser frequency $\omega_L$
and the value of the effective Rabi frequency $\Omega_{eff}$.

In the moving coordinate we may write the time dependent
Schr\"odinger equation as~\cite{kn:saif2}
\begin{equation}
i \hbar\dot\chi 
=\left\{\frac{p^2}{2M} + Mz(g-\lambda\omega^2\sin\omega t) 
+ V_0 e^{-\kappa z}\right\}\chi\;,
\label{schr1}\end{equation}
where $z=\tilde z- \lambda\sin\omega t$. Here we are 
considering $V_0\equiv\hbar\Omega_{eff}/4$, 
the steepness $\kappa\equiv 2\omega_L/c$ and  the
modulation strength $\lambda\equiv c\epsilon/(2\omega_L)$.

\begin{figure}
\psfig{figure=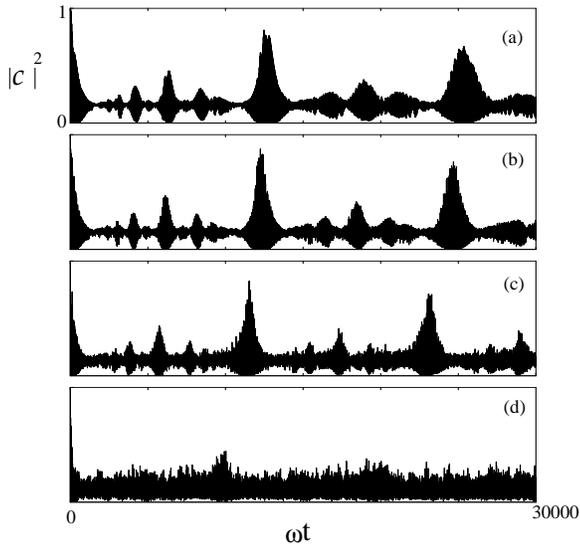,width=3.in}
\vspace{0.2truecm}
\caption{Revival phenomena in a
gravitational cavity in the absence and in the presence 
of periodic modulation: 
Autocorrelation of a Gaussian      
wave packet of a cesium atom as a function of time, prepared 
at $t=0$ with $\Delta z=0.28\mu m$, 
for $\lambda=0$ (a), $\lambda=0.56 \mu m$ (b),
$\lambda=1.13 \mu m$ (c), and $\lambda=2.26\mu m$ (d). The parameters are
$\omega=2\pi\times 0.93$KHz, $\Omega_{eff}=23.38$KHz, and $\kappa=0.57\mu m$. 
The average position of the 
wave packet in the gravitational cavity is $z_0=20.1\mu m$ 
which corresponds to the mean quantum number $n_0=176.16$.
}
\label{fg:reviv}
\end{figure}

A convenient way of obtaining insight into the 
influence of the acusto-optical external
driving force on the atomic dynamics is to
investigate the time evolution of
an atomic center-of-mass wave packet.
For this purpose we consider a Gaussian wave packet 
$\psi(0)=(2\pi\Delta z^2)^{-1/4}
\exp\{-(z-z_0)^2/(2\Delta z)^2\}\exp(-ip_0(z-z_0)/\hbar)$
at $t=0$ and propagate it in the gravitational cavity
for different modulation strength $\lambda$.
Here $z_0$ describes the average position, $p_0$ denotes
the average momentum of the wave packet and $\Delta z$ is the
spatial uncertainty. 
In Fig.~\ref{fg:reviv}  characteristic time dependences of the
autocorrelation function ${\cal C}(t)\equiv\langle\psi(0)|\psi(t)\rangle$
of the wave packet are shown for $z_0=20.1\mu m$, $\Delta z=0.28\mu m$ and $p_0=0$. 
Without external periodic perturbation (uppermost figure)
the well known scenario of revivals and fractional revivals~\cite{kn:aver} 
is clearly apparent. In the presence of 
a sufficiently weak external periodic
driving force the revivals and fractional revivals are still observable.
However, the revivals decrease in magnitude and the revival
time exhibits a pronounced
dependence on the external driving force. 
We find that if the modulation strength $\lambda$ exceeds
$0.25$ these revival phenomena disappear. 

In order to obtain insight into the quantitative dependence of the
revival time of an atomic wave packet on the external driving force
let us first concentrate on the case of a weak, off-resonant
periodic driving which can be treated perturbatively.
The revival time $T$ of a wave packet~\cite{kn:aver} which is centered
around mean energy $E_{n_0}$ is determined by the spectrum
of the nearby discrete quasi-energies ${\cal E}_n$ of the Hamiltonian
appearing in Eq.~(\ref{schr1}), i.e.
\begin{eqnarray}
T &=& \frac{4\pi \hbar}{\mid
\frac{d^2{\cal E}_n}{d n^2}
\mid_{n=n_0}}.
\label{T}
\end{eqnarray}
As long as the influence of the external, periodic force
can be described perturbatively these quasi energies are approximately
given by ${\cal E}_n = E_{n} + \delta E_n$ with the quadratic
Stark shift
\begin{equation} 
\delta E_n=\frac{(M\lambda\omega^2)^2}{4}\sum_{m=-\infty}^{\infty} 
\frac{2|\langle n+m|z|n\rangle|^2 (E_{n}-E_{n+m})}{(E_{n}-E_{n+m})^2-{(\hbar\omega)}^2}, 
\label{stark}
\end{equation} 
and with $E_{n}$ denoting the unperturbed eigenenergies of the
Hamiltonian of Eq.~(\ref{schr1}) with $\lambda=0$.
For large values of the quantum number $n$ asymptotic expressions
are obtainable for $E_{n}$ by using semiclassical quantization,
namely, $E_{n}\equiv\frac{1}{2}(Mg^2)^{1/3}[3\pi \hbar(n+3/4)]^{2/3}$.
Asymptotically valid expressions for the relevant matrix elements
$\langle n+m|z|n\rangle$ can be derived with the help of the
Bohr correspondence principle~\cite{kn:lind} 
in which these matrix elements
are related directly to the Fourier coefficients of the classical,
unperturbed motion of the atom in the triangular potential well,
namely
\begin{eqnarray}
\langle n+m| z | n\rangle \equiv
-\frac{2E_{n}}{M\pi^2m^2g}\left(1+\frac{m}{3(n+3/4)}\right).
\end{eqnarray}
This way all quantities appearing in Eq.~(\ref{T})
can be evaluated from the corresponding unperturbed, classical dynamics.
Inserting these asymptotic expressions into Eq.~(\ref{stark})
the semiclassical approximation for the Stark shift
\begin{equation} 
\delta E_n=-(\omega^2\lambda/g)^2\frac{3E_{n}}{\pi^4}\sum_{m=1}^{\infty}m^{-4}\left( 
\frac{1+\frac{2/3}{1-(\Omega_n/m)^2}}{1-(\Omega_n/m)^2}\right) 
\label{semi}
\end{equation} 
is obtained with $\Omega_n=3(n+3/4)\hbar\omega/(2E_{n})$
and with $m$ being integer.
Thereby the quadratic Stark shift is expressed as a sum of contributions
of virtual oscillators $m$ which describes the influence of the
external periodic force on the bound motion of the atom in the
triangular potential well.
As the effective coupling strength of these
virtual oscillators to the external field decreases rapidly
with increasing $m$
the dominant contribution to the quadratic Stark shift is given by
the first few terms in the sum of Eq.~(\ref{semi}).
Eq.~(\ref{semi}) describes the global frequency dependence of the
Stark shift and is valid in the perturbative regime for small enough
values of the driving strength $\lambda$. It breaks down close to 
resonances at which $(\Omega_n/m)^2=1$.
Inserting Eq.~(\ref{semi}) into Eq.~(\ref{T})
we obtain the revival time 
\begin{eqnarray} 
T_{\lambda}
&&=T_0\left[1-(\omega^2\lambda/g)^2\frac{3}{\pi^4}\sum_{m=1}^{\infty}
\frac{m^{-4}}{E''_{n_0}} \right. \nonumber\\
&&\left. \left. \frac{d^2}{d n^2}\left\{E_n\left( 
\frac{1+\frac{2/3}{1-(\Omega_n/m)^2}}{1-(\Omega_n/m)^2}\right)\right|_{n=n_0}    
\right\}\right]
\label{revpert}
\end{eqnarray} 
where $T_0$ is the revival time 
in the absence of the external modulation.

Alternatively one can calculate the revival time with the help
of semiclassical secular theory which is expected to yield
a particularly good approximation close to a resonance. 
In the vicinity of the $N$-th primary resonance the
classical  Hamiltonian corresponding to Eq.~(\ref{schr1})
can be expressed by action and angle variables ($I,\varphi$)~\cite{kn:saif2,kn:lieb} as 
\begin{eqnarray}
H=\frac{H''}{2}(I-I_0)^2+H'(I-I_0)\nonumber\\
+ H_0(I_0) +\lambda V\sin(N\varphi-\omega t)\;.
\label{eq:reshsc}
\end{eqnarray}
Here $I_0$ is the classical action associated with the initial
condition and
$V=(Mg)(I_0/I_N)^{2/3}$. The classical action at the 
center of the $N$-th primary resonance is
denoted by $I_N$ . 
Moreover $H'$ and $H''$ are the first and second derivatives of 
energy with respect to action calculated at $I_0$. Similarly
$H_0(I_0)$ is the energy of the unperturbed system for the initial
action $I_0$.
In Eq.~(\ref{eq:reshsc}) we have averaged 
out the fast oscillating terms so that $H(I-I_0)$ represents
an integrable one-degree-of-freedom physical system.

Introducing the transformation $N\varphi-\omega t=2\theta+\pi/2$
and quantizing the dynamics around the $N$-th resonance 
by using $I-I_0=\frac{\hbar}{i}\frac{\partial}{\partial\varphi}
=\frac{N\hbar}{2i}\frac{\partial}{\partial\theta}$~\cite{kn:berry},
the time independent Schr\"odinger equation becomes
\begin{eqnarray}
\left[-\frac{N^2H''{\hbar}^2}{8}\frac{\partial^2}
{\partial\theta^2}+\frac{\hbar}{2i}(NH'-1)
\frac{\partial}{\partial\theta} + H_0(I_0)\right.\nonumber\\
\left. +\lambda V\cos 2\theta\right]\psi={\cal E}_n\psi\;.
\label{eq:schham}
\end{eqnarray}
Thus the quasi energies ${\cal E}_n$ are determined 
by Eq.~(\ref{eq:schham}).
Here $\psi(\theta)$ has to fulfill the periodic boundary condition.
It is straightforward to write Eq.~(\ref{eq:schham}) in the form of 
a Mathieu equation by substituting 
$\psi=\phi\exp\left(-2i(NH'-1)\theta/(N^2H''\hbar)\right)$,
namely
\begin{equation}
\left[\frac{\partial^2}{\partial\theta^2} +a -2q\cos 2\theta\right]\phi=0\;,
\label{eq:math}
\end{equation}
with
\begin{eqnarray}
a&=&\frac{8}{N^2H''(I_0){\hbar}^2}\left[{\cal E}_n-H_0(I_0)
+ \frac{(NH'(I_0)-1)^2}{2N^2H''(I_0)}
\right]\;,\\
q&=&\frac{4\lambda V}{N^2H''(I_0){\hbar}^2}\;.
\end{eqnarray}
The quasi-energy eigenvalues ${\cal E}_n$ are determined 
by the solutions of Eq.~(\ref{eq:math}) and the 
requirement that $\phi(\theta+\pi)=\phi(\theta)$.
The $\pi$-periodic solutions of Eq.~(\ref{eq:math}) correspond to 
even functions of the Mathieu equation whose corresponding
eigenvalues are real~\cite{kn:abra}. 
These solutions are $\phi_{\nu}(\theta)=e^{i\nu\theta}P_{\nu}(\theta)$
where $P_{\nu}(\theta)$ is the even order Mathieu function.
In order to obtain a $2\pi$-periodic solution in $\varphi$-coordinate
we require the coefficient of $\varphi$ to be equivalent to  
an integer number. This requirement
provides us the oppertunity to find the value for 
the index $\nu$ of the Mathieu functions as 
\begin{eqnarray}
\nu=\frac{2}{N\hbar}\left[I-4I_0
+3I_0\left(\frac{I_0}{I_N}\right)^{1/3}\right]\;,
\label{eq:denu} 
\end{eqnarray}
where $I=(n+3/4)\hbar$. 
The quasi energy of the system is finally given by
\begin{equation}
{\cal E}_{n}=\frac{N^2H''{\hbar}^2}{8}a_{\nu(n)}(q)
-\frac{(NH'-1)^2}{2N^2H''}
+H_0(I_0)\;,
\label{eq:qen}
\end{equation}
with the Mathieu characteristic parameter
$a_{\nu(n)}(q)$.

From the quasi-energies of Eq.~(\ref{eq:qen})
we can determine the revival time in the presence of the external time 
dependent field. Considering $q<1$ 
we expand the Mathieu characteristic parameter $a_{\nu(n)}(q)$
up to $q^2$. We simplify our result by noting that
$H''=-(Mg^2)^{1/3}(\pi/9I_0^2)^{2/3}$ and $N=(\omega^3/Mg^2)^{1/3}(3I_N/\pi^2)^{1/3}$. 
Thus we finally obtain
\begin{equation}
T_{\lambda}=T_0\left[ 1-\frac{1}{2}\left\{\frac{8\lambda}
{{\hbar}^2} E_N \left( \frac{I_0}{I_N}\right)^2\right\}^{2}
\frac{3\nu^2+1}{(\nu^2-1)^3}\right].
\label{eq:rtspt}
\end{equation}
On substituting the value of $\nu$, calculated at $I=I_0$, in
Eq.~(\ref{eq:rtspt}) we get
\begin{equation}
T_{\lambda}=
T_0\left[ 1-\frac{1}{8}\left\{
    \frac{M\lambda g}{E_{n_0}} \right\}^{2}
    \frac{3(1-r)^2+a^2}{((1-r)^2-a^2)^3}\right]
\label{eq:rtsptn1} 
\end{equation}
where $r\equiv(E_N/E_{n_0})^{1/2}$ and $a\equiv r^2 \hbar\omega/4E_{n_0}$.
If the initial energy is large, {\it i.e.} $E_{n_0}\gg \hbar\omega$, we 
may consider $a^2$ much smaller than $(1-r)^2$, which leads to
\begin{equation}
T_{\lambda}=T_0\left[ 1-\frac{3}{8}\left\{ 
\frac{M\lambda g}{E_{n_0}} \right\}^{2} 
\frac{1}{(1-r)^4}\right].
\label{eq:rtsptn} 
\end{equation}

\begin{figure}
\psfig{figure=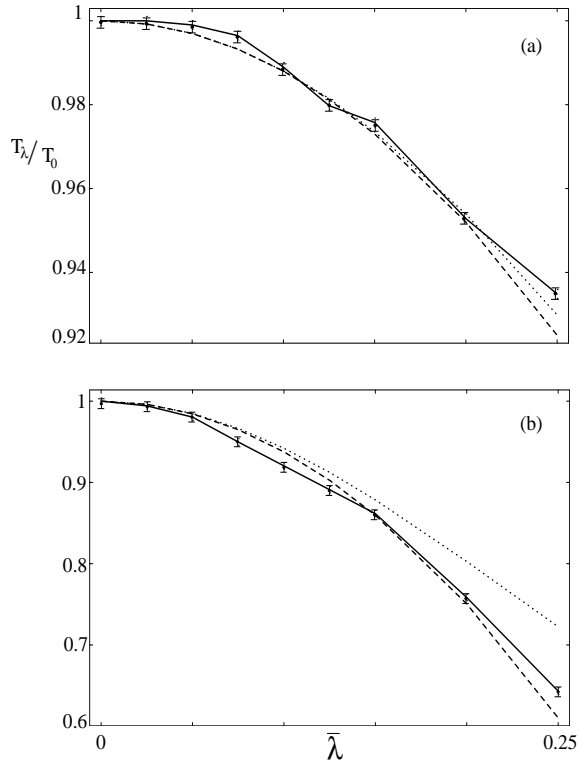,width=3.in}
\vspace{0.3truecm}
\caption{Revival times $T_{\lambda}$ as obtained from
exact numerical calculations of the time dependent 
Schr\"odinger equation (\ref{schr1}) (solid lines), 
and from analytical results based on
Eqs.~(\ref{revpert}) (dotted lines) and 
~(\ref{eq:rtspt}) (dashed line): 
Initial conditions are (a) $z_0=29.8\mu m$, 
$r=0.87$ and (b)  $z_0=20.1\mu m$, 
$r=0.77$. The other parameters are the same as in Fig.~\ref{fg:reviv}.
Here we have $\bar\lambda=\lambda\omega^2/g$.
The error bars indicate the uncertainty
in determining the revival times from the numerical
data. In all cases this uncertainty is less than 2 percent.
}
\label{fg:plot}
\end{figure}

The analytical results of Eqs.~(\ref{revpert}) and (\ref{eq:rtsptn})
are main results of this paper. They explain in a simple
way the quantitative dependence of the revival time $T_{\lambda}$
on the characteristic parameters of the problem, namely
the driving frequency $\omega$ and the driving amplitude $\lambda$.
In order to access the accuracy of these perturbative results
we calculate the revival time $T_{\lambda}$ by integrating the Schr\"odinger 
equation Eq.~(\ref{schr1}) numerically, and compare it with the analytically
obtained results of Eqs.~(\ref{revpert}) and~(\ref{eq:rtsptn}). 
For this comparison we have considered
two different initial conditions of the atomic
center-of-mass wave packet above 
the surface of the atomic mirror. In Fig.~\ref{fg:plot}a 
$z_0=29.8\mu m$ which corresponds to a state with mean principle 
quantum number $n_0=322.51$.
In Fig.~\ref{fg:plot}b $z_0=20.1\mu m$ which implies $n_0=176.16$.
In both cases $p_0=0$.
The first initial condition lies further away from
the center of the corresponding primary resonance as compared to the
second one. As expected, in Fig.~\ref{fg:plot}a the
revival time obtained by means of 
second order perturbation theory agrees well with the numerical results
whereas close to resonance it deviates from the numerical result in
Fig.~\ref{fg:plot}b. However, the revival time obtained 
with the help of semiclassical secular theory agrees well
with the numerical results in both cases. 

Numerically we find that the change in revival time depends
quadratically on the strength of external modulation $\lambda$
as predicted from Eqs.~(\ref{revpert}) and 
(\ref{eq:rtsptn}) and displayed in Fig.~\ref{fg:plot}.
The change of the revival time is smaller in the case depicted in
Fig.~\ref{fg:plot}(a) than in the case shown in Fig.~\ref{fg:plot}(b). 
This can be understood from our analytical result:
In case (a) of Fig.~\ref{fg:plot} our 
chosen initial condition $z_0=29.8\mu m$
has a higher energy $E_{n_0}$ than in case (b) 
where $z_0=20.1\mu m$.
Since the change in revival time has inverse dependence
on the square of the energy $E_{n_0}$, as a result, we observe
a smaller change in revival time for case (a) as compared to case (b).

We have demonstrated that the dynamics of a material wave packet
in a periodically driven gravitational cavity exhibits quantum 
mechanical revivals as long as the driving strength of the 
periodic force does not exceed a critical value. First
results on the quantitative dependence of the revival time
on the characteristic parameters of the problem, namely 
the driving frequency and the driving strength have been 
presented. It has been shown that this dependence
can be understood quantitatively in a satisfactory way by
using semiclassical perturbation theory. 
In view of recent experimental~\cite{kn:sten} developments 
the presented quantitative predictions 
should be accessible to experimental observation.

We thank I.\ Bialynicki-Birula, M.\ Fortunato, 
M.\ El Ghafar, R.\ Grimm, P.\ T\"orm\"a,
and A.\ Zeiler
for many fruitful discussions.

\newpage

\begin{references}
\bibitem{kn:benv1} E.J. Galvez, B.E. Sauer, L. Moorman, P.M. Koch and D. Richards,
                   Phys. Rev. Lett. {\bf 61}, 2011 (1988);
                   M. El Ghafar, P. T\"orm\"a, V. Savichev, E. Mayr,
                   A. Zeiler and W. P. Schleich, Phys. Rev. Lett. {\bf 78},
                   4181 (1997); For a review 
                   see M. G. Raizen, in {\it Advances in Atomic,
                   Molecular, and Optical Physics} Vol. 41, p. 41
                   ed. B. Bederson and H. Walther 
                   (Academic Press, New York 1999).
\bibitem{kn:hogg} T. Hogg and B. A. Huberman,  Phys. Rev. Lett. {\bf 48},
                  711 (1982).
\bibitem{kn:haak} F. Haake, {\it Quantum Signatures of Chaos}, 
                  (Springer, Berlin 1992).
\bibitem{kn:bres} J. K. Breslin, C.A. Holmes, and G.J. Milburn,
                  Phys. Rev. A {\bf 56}, 3022 (1997).  
\bibitem{kn:toms} S. Tomsovic and J. Lefebvre,  
                  Phys. Rev. Lett. {\bf 79}, 3629 (1997).  
\bibitem{kn:mlyn}  For a review of the field of atom optics see the
                   special issue, E. Arimondo and H.A. Bachor, (eds.), 
                   Quantum Semicl. Opt.
                  {\bf 8}, 495 (1996).
\bibitem{kn:wallis} H. Wallis, J. Dalibard, and C. Cohen-Tannoudji, Appl. Phys. B
                 {\bf 54}, 407 (1992).
\bibitem{kn:sten} A. Steane, P. Szriftgiser,
                 P. Desbiolles and J. Dalibard, Phys. Rev. Lett. {\bf 74}, 4972
                 (1995).
\bibitem{kn:saif1} F. Saif, I. Bialynicki-Birula, M. Fortunato, 
                 and W.P. Schleich, Phys. Rev. A 4779, {\bf 58} 1998. 
\bibitem{kn:fermi} E. Fermi, Phys. Rev. {\bf 75}, 1169 (1949).
\bibitem{kn:saif2} For a comprehensive study of revival phenomena 
                   in driven systems see e.g.
                   F. Saif, Ph.D thesis, (Universit\"at Ulm, 1998).
\bibitem{kn:aver} I. Sh. Averbukh and N.F. Perel'man, Phys. Lett. A {\bf 139}, 
                  449 (1989). 
\bibitem{kn:lieb} A.J. Lichtenberg and M.A. Lieberman, {\it Regular and Stochastic
                  Motion}, (Springer, Berlin, 1983) and references therein.
\bibitem{kn:lind} L.D. Landau and E.M. Lifschitz, {\it Quantenmechanik},
                  (Akademie-Verlag, Berlin, 1971).
\bibitem{kn:berry} M. V. Berry, Philos. Trans. R. Soc. London, 
                   Ser. B {\bf 287}, 237 (1977). 
\bibitem{kn:abra} M. Abramowitz and I.A. Stegun, 
                  {\it Handbook of Mathematical Functions}, 
                  (Dover, New York, 1992). 
\end{references}
\end{document}